\documentclass[a4paper,twoside, twocolumn]{cpc-hepnp}
\usepackage[english]{babel}
\usepackage{lineno,hyperref, xspace,fancyhdr}
\modulolinenumbers[1]

\usepackage{makeidx}
\usepackage{color}
\usepackage{listings}
\usepackage{bm}
\usepackage{url}
\usepackage{booktabs}
\usepackage{amsmath}
\usepackage{paralist}
\usepackage{amssymb}
\usepackage{algorithm}
\usepackage{algorithmic}
\usepackage{upgreek}
\usepackage{textcomp}
\usepackage{multicol}
\usepackage{graphicx}
\usepackage{subfigure}
\usepackage{multirow}
\usepackage{threeparttable}
\usepackage{tabularx}
\usepackage{array}

\newcommand{\PreserveBackslash}[1]{\let\temp=\\#1\let\\=\temp}
\newcolumntype{C}[1]{>{\PreserveBackslash\centering}p{#1}}
\newcolumntype{R}[1]{>{\PreserveBackslash\raggedleft}p{#1}}
\newcolumntype{L}[1]{>{\PreserveBackslash\raggedright}p{#1}}

\usepackage{soul}
\soulregister\cite7
\soulregister\ref7
\soulregister\pageref7

\begin{document}

\title{Enhanced search sensitivity to the double beta decay of $^{136}$Xe to excited states with topological signatures}

\maketitle

\author{%
	\quad\quad\quad\quad\quad\quad\quad\quad\quad\quad\quad\quad Chen Xie$^{1}$
	\quad Kaixiang Ni$^{1}$
	\quad Ke Han$^{1,\dagger}$
	\quad Shaobo Wang$^{1,2,\ddagger}$
}

\address{%
	~\\
	$^1$ Institute of Particle and Nuclear Physics and School of Physics and Astronomy,
	Shanghai Jiao Tong University,\\ Shanghai Laboratory for Particle Physics and Cosmology, Shanghai {\rm 200240}, China\\
	$^2$ ParisTech Elite Institute of Technology,
	Shanghai Jiao Tong University, Shanghai  {\rm 200240}, China\\
}

\begin{abstract}
Double beta decay of $^{136}$Xe to excited states of $^{136}$Ba (DBD-ES) has not yet been discovered experimentally.
The experimental signature of such decays, one or two gamma rays following the beta signals, can be identified more effectively in a gaseous detector with the help of topological signatures.
We have investigated key parameters of particle trajectories of DBD-ES with Monte Carlo simulation data of the proposed PandaX-III detector as an example.
The background rates can be reduced by about one order of magnitude while keeping more than half of signals with topological analysis.
The estimated half-life sensitivity of DBD-ES can be improved by 1.8 times to  4.1$\times$10$^{23}$ yr (90\% CL).
Similarly, the half-life sensitivity of neutrinoless double beta decay of $^{136}$Xe to excited states of $^{136}$Ba can be improved by a factor of 4.8 with topological signatures.
\end{abstract}



\footnotetext[0]{$\dagger$ Corresponding author: ke.han@sjtu.edu.cn}%
\footnotetext[0]{$\ddagger$ Corresponding author: shaobo.wang@sjtu.edu.cn}%

\begin{multicols}{2}

\section{Introduction}

Double beta decay~(DBD) is a rare nuclear process in which two beta decays happen simultaneously in a nucleus~\cite{2nbb}.
The decay to the ground state of its daughter, which is commonly referred to as DBD for short,  has been observed in 11 isotopes, including $^{136}$Xe, $^{130}$Te, and $^{76}$Ge.
Most of the half-lives are in the range of $10^{19}$ to $10^{24}$ years.
Double beta decay to the excited states (DBD-ES), due to the smaller branching ratio and longer half-lives, has only been observed in a couple of isotopes.
DBD of $^{136}$Xe to the ground states of $^{136}$Ba was discovered by EXO-200~\cite{EXO-200:2nbb} and then confirmed by KamLAND-Zen~\cite{KamLAND-Zen:2nbb}.
The two collaborations have searched for DBD-ES of $^{136}$Xe with null results so far.
If DBD-ES is observed, the comparison between half-lives of DBD and DBD-ES can provide critical input to evaluate the nuclear matrix elements (NME)~\cite{NME} of the decay and help to understand the rare nuclear process in general.

The hypothetical DBD process without neutrino released, which is called neutrinoless DBD (NLDBD), is of great importance for nuclear and particle physics.
NLDBD is a lepton-number-violating process~\cite{LeptionViolating} and would also prove the Majorana nature of neutrinos~\cite{Majorana}.
NLDBD to ground states are actively being searched in different candidate isotopes, and the established half-life limits can be as long as $10^{26}$ years.
NLDBD to excited states (NLDBD-ES) is also possible but with even longer half-lives compared to the decay to ground states.

The $^{136}$Xe decay to the $ 0_{1}^{+}$ excited state of $^{136}$Ba with the energy released of 878.8~keV is the dominant one among all possible (NL)DBD-ES channels. The energies of the subsequent de-excitation $\gamma$ rays are 760.5~keV and 818.5~keV respectively.
Hence, we focus our studies on the decay to the $0_{1}^{+}$ excited state.
In the NLDBD-ES case, two electrons carry away the total energy.
While in DBD-ES, the two-electron energy spectrum is a continuous one with an endpoint at 878.8~keV.

In experimental searches, the de-excitation $\gamma$ ray is essential for enhancing the signal-to-noise ratio.
Coincidence between electron-pair and de-excitation $\gamma$ rays can be found in multiple modules of a detector array or different parts of a large monolithic detector.
Gaseous detectors can measure the detailed energy deposition along particle trajectories and identify the electron-pairs and $\gamma$ rays by the trajectory characteristics.
We argue that the topological analysis on particle tracks can enhance the (NL)DBD-ES search sensitivity.

Herein, the proposed PandaX-III~\cite{PandaX-III CDR} detector is taken as an example of a high-pressure gaseous time projection chamber (TPC)~\cite{TPC}.
The objective of PandaX-III is to build a detector with 140 kg of 90\% $^{136}$Xe-enriched xenon to search for NLDBD of $^{136}$Xe in the China Jin-Ping underground Laboratory (CJPL).
The active volume is 1.6~m in diameter and 1.2~m in length, and it will be operated at 10 bar pressure.
Xenon gas will be mixed with approximately 1\% Trimethylamine to suppress the scintillation light of xenon and diffusion effects of electron drift~\cite{TMA1, TMA2}.
Subsequently, only ionization signals are readout and amplified with the Micromegas module (MM)~\cite{MMApplication}.
The target energy resolution at $ Q_{NLDBD}=2457.8$~keV is 3\% full width at half maximum~(FWHM)~\cite{EnergyResolution}.

\begin{figure}[H]
	\centering
\includegraphics[width=\columnwidth]{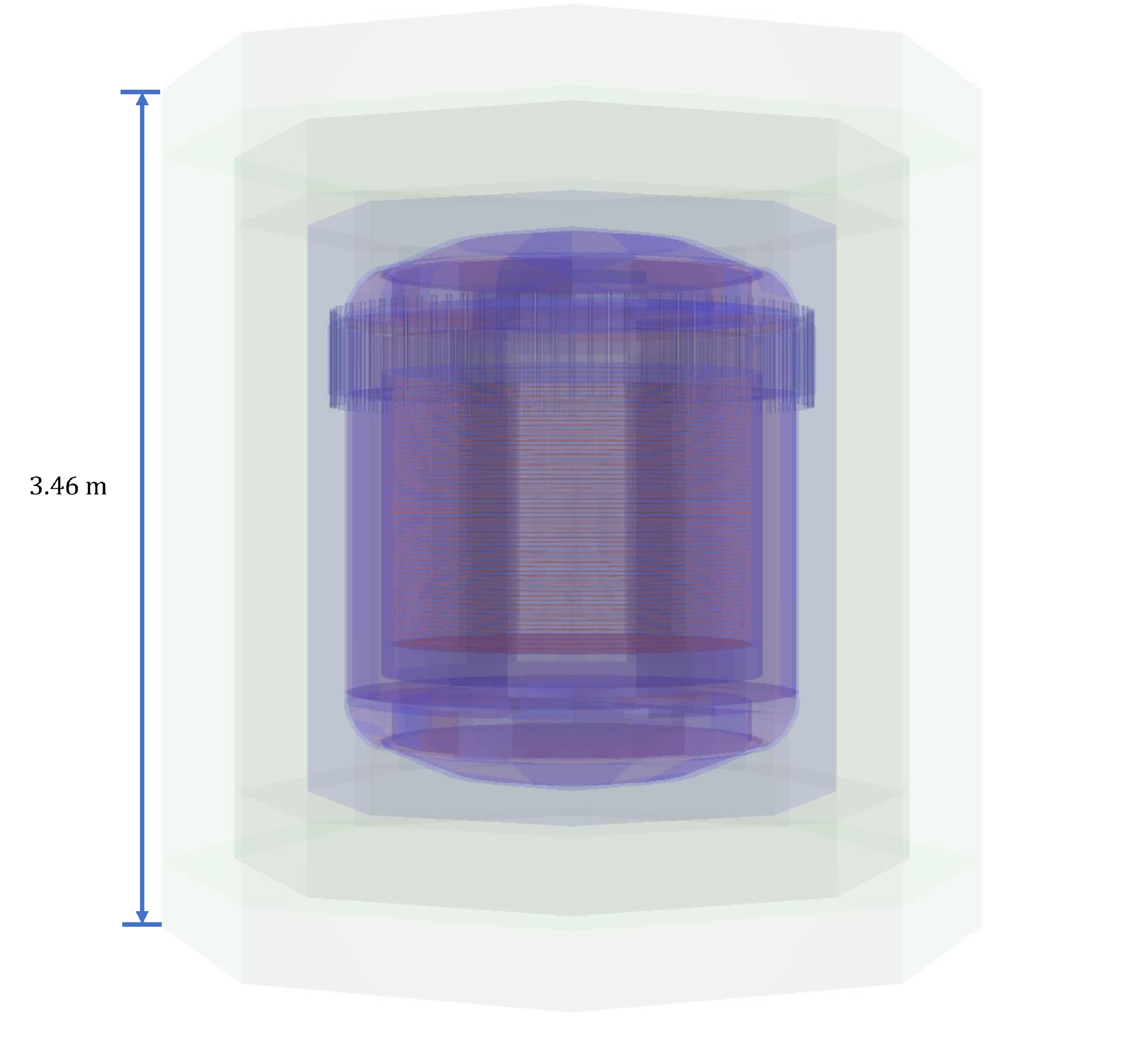}
	\caption{Illustration of the PandaX-III detector reconstructed in MC simulation. The most prominent features include two outer shielding layers (light green), an SS vessel (light blue), and copper pieces (orange--red). }
	\label{TPC structure}
\end{figure}

\section{Detector simulation with Geant4}
\label{sc:Geant4}

Our Monte Carlo simulation is based on the detector geometry of the current conceptual design of the PandaX-III experiment~\cite{Pandax-iii}.
The detector is enveloped in a stainless steel (SS) cylindrical vessel (Figure~\ref{TPC structure}).
The vessel barrel and end-caps are 1 and 1.8~cm thick, respectively.
 The barrel is welded to the bottom torispherical cap, and two flanges connect the top torispherical cap to the barrel. The total SS mass is approximately 2.5~t.

The detector comprises a single-ended design with a readout plane on top and a cathode at the bottom.
A cylindrical field cage connecting the two parts consists of a 5-cm-thick acrylic barrel that weighs 723~kg.
Copper rings are embedded in the acrylic barrel for electric field shaping.
The charge readout plane with 52 $ 20\times20 $ cm$ ^{2} $ MMs covers most of the TPC's active volume.
The active area of each MM is divided into 3~mm diamond-shaped pads that are connected in horizontal and vertical directions and read as 128 X- and Y- strips, respectively.
Using the timing information from the Z-direction, PandaX-III records two two-dimensional (2D) tracks in the X--Z and Y--Z planes instead of one three-dimensional track.
The total energy of the physical trajectory is shared between two 2D tracks.
More details on the Micromegas readout can be found in~\cite{Pandax-iii MMs}.

Several layers of shielding are implemented to reduce the gamma background from the SS vessel and lab environment.
A layer of 12.5-cm-thick ultra-pure oxygen-free copper is inserted between the field cage and SS vessels.
Two torispherical copper shielding blocks above and below the TPC are also added.
The total mass of the copper liner reaches 22.6~t.
One layer of lead and one layer of high-density polyethylene (HDPE) are placed outside the vessel.
The thickness of these two shielding layers is 30 cm each.
\begin{figure}[H]
	\centering
	\includegraphics[width=\linewidth]{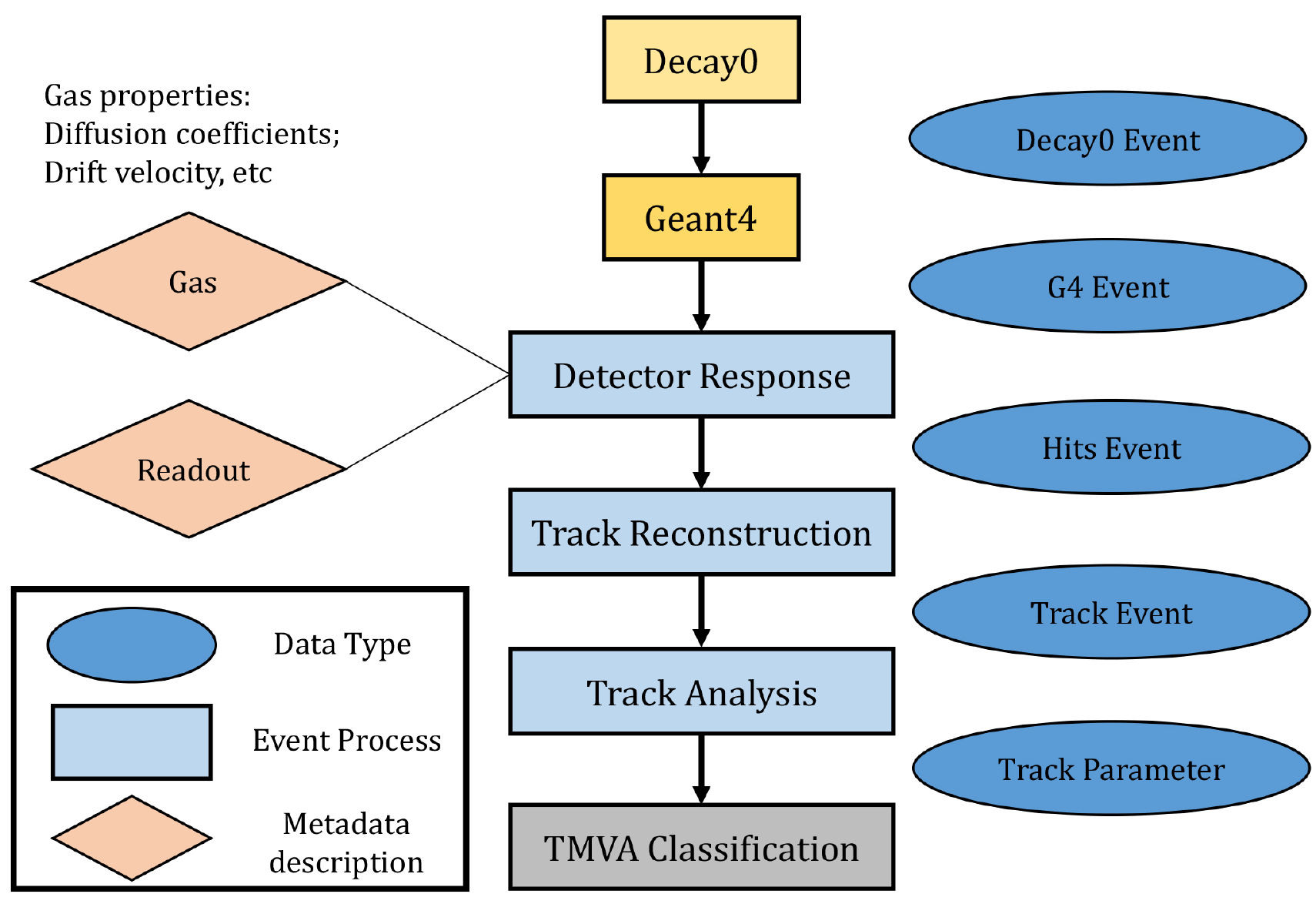}
	\caption{Work flow of simulation and analysis in six stages, shown in rectangles from top to bottom.
		Detector responses, such as gas medium properties and charge readout schemes, are introduced in the third stage.
		The blue ovals on the right denote data types transmitted between adjacent stages.}
	\label{work flow}
\end{figure}

Figure~\ref{work flow} illustrates the complete workflow for simulating and analyzing (NL)DBD-ES signals in six stages. In the first stage, the energy and angular distribution of particles emitted from (NL)DBD-ES events are generated via the Decay0 package~\cite{Decay0}.
Simulated signal events are used as input for Geant4~\cite{Geant4} with the above described geometry.
Simulation of background events starts from the second stage.
For the PandaX-III detector setup, we simulate background contributions from major sources, including the bulk contamination of $^{238}$U and $^{232}$Th in Micromegas, copper liner, SS vessel, and acrylic field cage.
For the copper liner, $^{60}$Co and $^{40}$K contaminations are also considered.
The contamination levels used in the simulation are shown in Table~\ref{table:activity}.
We also consider the Micromegas surface radioactivity, which has an upper limit of 45 and 14~nBq/cm$^{2}$ for $^{238}$U and $^{232}$Th respectively~\cite{MMRatio}.

Large sets of signal and background data are generated and simulated to obtain sufficient statistics on training and testing samples in the efficiency studies later.
We simulate 10 and 20 million DBD-ES and NLDBD-ES events, respectively.
The simulated background events from different radioisotopes in different detector components are weighted by contamination levels and mass of the component.
For example, 6.52$\times$10$^{8}$ $^{238}$U chain, 1.74$\times$10$^{8}$ $^{232}$Th chain, 1.74$\times$10$^{9}$ $^{60}$Co, and 2.00$\times$10$^{10}$ $^{40}$K events originating from copper liner are simulated.
Moreover, we generate 3.48$\times$10$^{7}$ (1.08$\times$10$^{7}$) $^{238}$U ($^{232}$Th) events from Micromegas and 3.80$\times$10$^{8}$ (1.24$\times$10$^{8}$) $^{238}$U ($^{232}$Th) events from the acrylic field cage.
Due to effective shielding of the copper liner, the efficiency of $\gamma$ rays from the SS vessel reaching the active volume of the detector is extremely low, and the effect of contamination from the SS vessel is scaled from that of copper liner by the radioactive levels and shielding effects.

\begin{table}[H]
	\centering
	\renewcommand\arraystretch{1.5}
	\resizebox{\columnwidth}{!}
	{
		\begin{tabular}{cccccc}
			\toprule[0.2mm]
			 \multirow{2}{*}{Component}	 &\multirow{2}{*}{Material}	 & \multicolumn{4}{c}{Activity ($\mu$Bq/kg)}  \\
                              &   &	$^{238}$U	& $^{232}$Th	& $^{60}$Co &  $^{40}$K \\
			\midrule[0.1mm]
			{Liner} &{Copper~\cite{copperRatioUTh,copperRatioCoK}  } &	0.75	&	0.20	&	2	& 23
			\\
			{Field Cage} &{Acrylic~\cite{acrylicRatio}} &	13.68	&	4.48	&	-	& -
			\\
            {Vessel} & {Stainless Steel~\cite{SSRatio} } &	500	&	320	& - &	-
			\\
			\bottomrule[0.2mm]
		\end{tabular}
	}
		\caption{Bulk radioactivities of isotopes  for major components considered in the simulation. }
	\label{table:activity}
\end{table}

\begin{figure}[H]
	\centering
	\includegraphics[width=\linewidth]{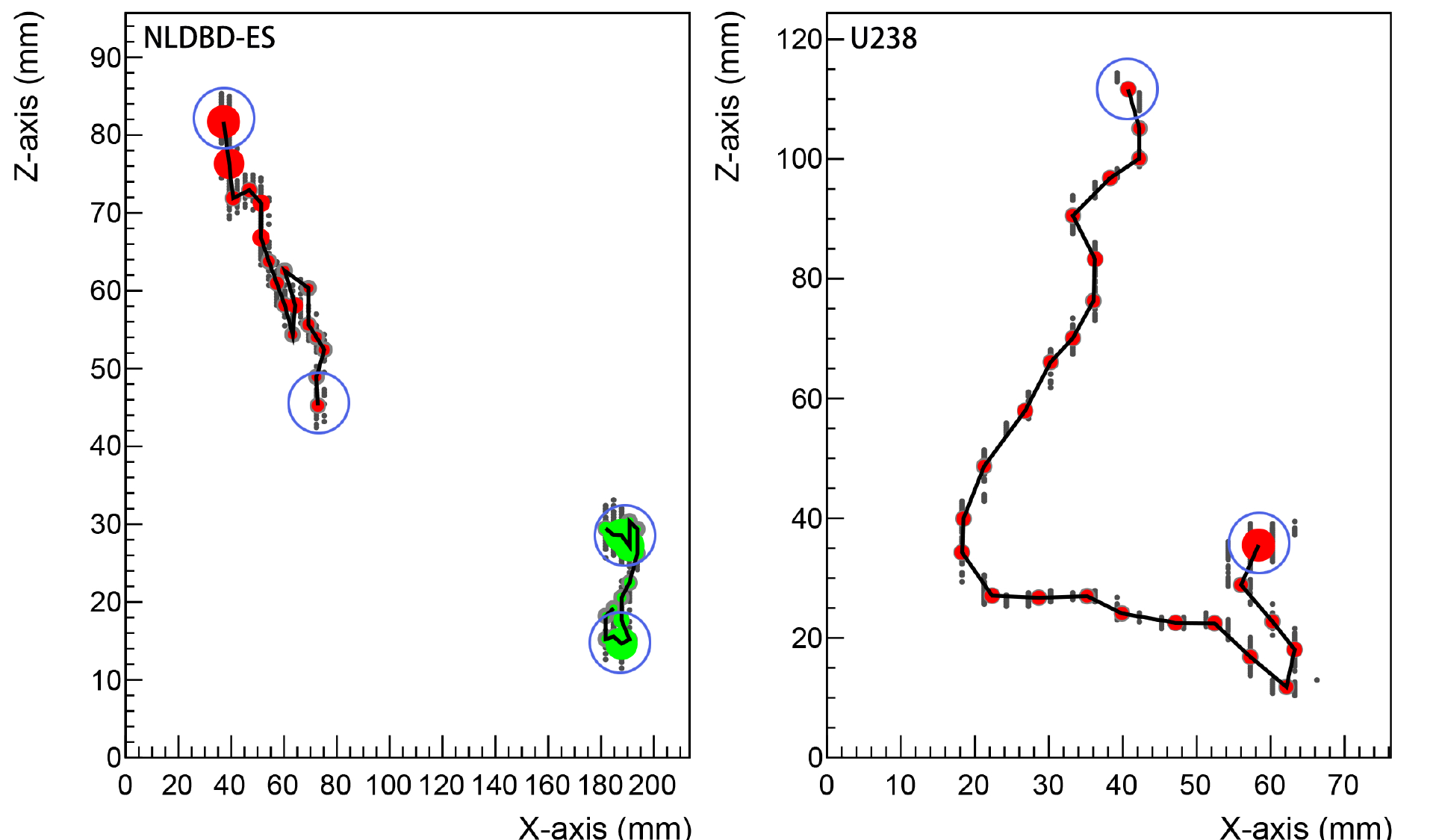}
	\caption{Reconstructed tracks of an example NLDBD-ES event (left) and a background event originating from Micromegas $^{238}$U (right).
	Only the X--Z projections of the tracks are shown.
	The total energy of both events is in the range of [1645.3, 1749.3]~keV.
}
	\label{NLDBD-ES_U238}
\end{figure}

The next three stages are implemented in REST~\cite{RESTInvention}, a software package developed for simulation and track reconstruction in TPC-based detectors.
Details on signal generation, simulation, reconstruction, and analysis in the framework can be found in~\cite{RESTInvention, REST}.
In the third stage, TPC responses, such as electron diffusion and energy smearing, are introduced.
Energy deposition is also grouped by the readout strips of MM and the effective coverage of the readout plane is considered.
In the fourth stage, energy and timing information from readout strips is used to reconstruct the particles' tracks.
Subsequently, vital parameters are extracted to characterize reconstructed tracks.
An example of track characteristics is blob energy, which describes the energy deposition at the end of a trajectory.
For an electron traveling in the medium, the energy loss per unit volume right before it stops is larger than that along the trajectory due to the Bragg peak and more meandering tracks at the end.
The energy loss within a spherical volume by the end of a track is called the blob energy, and the ratio of the smaller blob energy to a larger one is defined as \emph{blob ratio} or \emph{QR} for short.
The volume used to define QR is optimized for a specific detector setup.
For the PandaX-III readout scheme, QR is defined within a circle with a 4 mm radius in the X--Z or Y--Z plane.
The QR for a track of two electrons originating from (NL)DBD-ES is closer to one than that for the background events.
Figure~\ref{NLDBD-ES_U238} shows two example tracks from NLDBD-ES and a background event from the $^{238}$U chain.
The NLDBD-ES event (left) contains two electrons with a total energy of 878.8~keV (the green track) and one $\gamma$ ray with an energy of 818.5~keV (the red track).
The dual-electron track exhibits a distinct feature of two large blob energies.
The background event on the right can be seen with one small blob of energy at the top of the track.
Besides QR, other discriminating parameters will be introduced in Section~\ref{sc:bb2n_Ex} and \ref{sc:bb0n_Ex}, corresponding to the case of DBD-ES and NLDBD-ES, respectively.

\begin{figure*}[t]
	\centering
	\includegraphics[width=2\columnwidth]{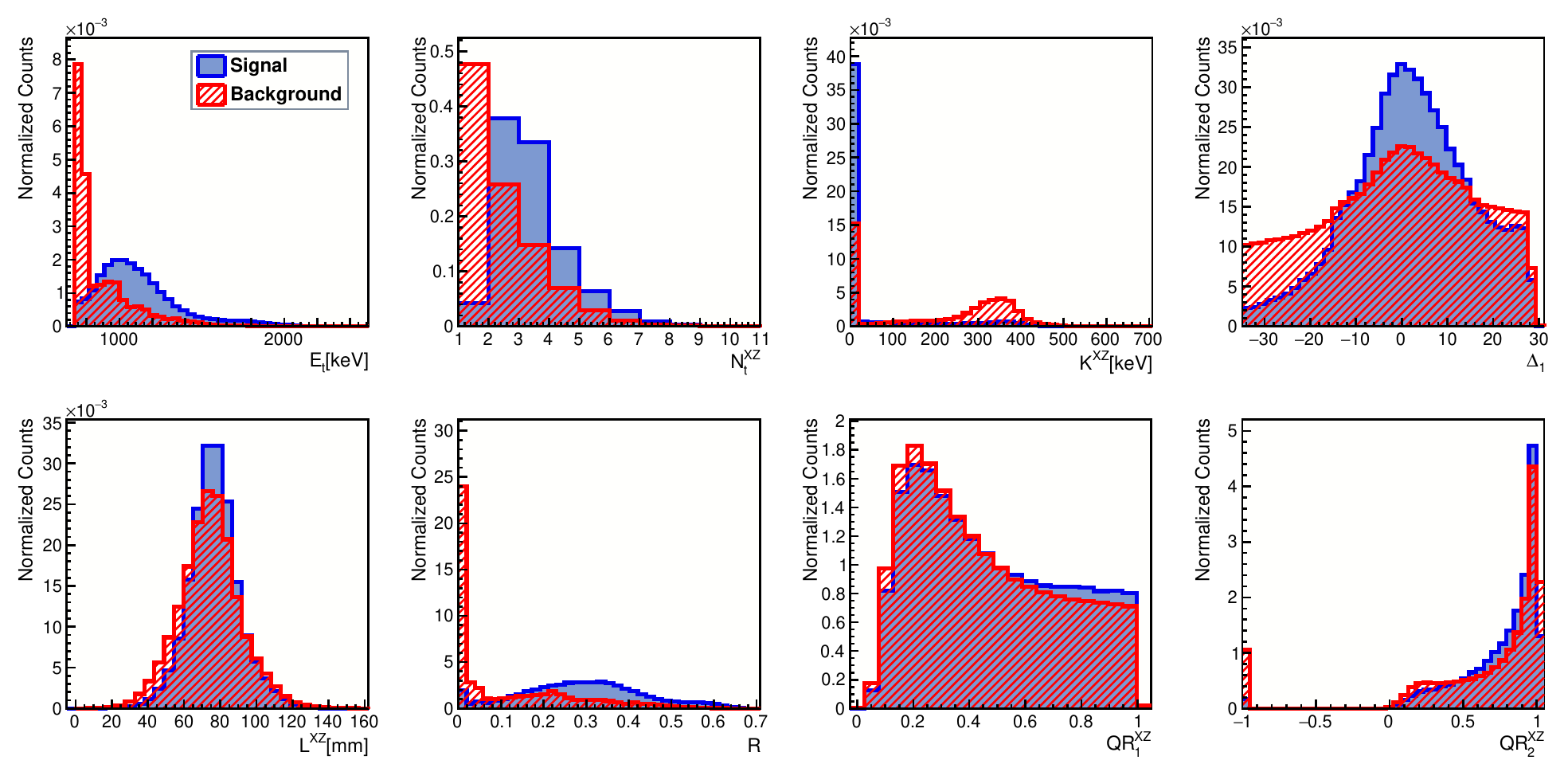}
	\caption{Distributions of the input topological parameters of TMVA in the MTE ROI [725.7, 789.0]~keV. Signal (DBD-ES) is shown in solid blue and background in hatched red.}
	\label{TMVA_DBD-ES}
\end{figure*}

Using toolkit for multivariate analysis (TMVA)~\cite{TMVA} in the ROOT framework~\cite{Root}, we classify events into signal or background with the distribution of the track parameters.
These backgrounds and signals are equally divided into random training and test datasets for TMVA classification.
TMVA ranks each test event with a discriminator variable between $-$1 and 1, denoting background- and signal-like events, respectively.
The classification is then compared with MC truth to calculate relevant efficiencies and subsequent significances.

Lastly, assuming no discovery, we can calculate the following equation to determine the lower limit for (NL)DBD-ES half-life at 90\% confidence level (C.L.).

\begin{equation}
    T_{1/2} > {\frac{\ln (2) N T \epsilon_{s}} {S_{up}} }
    \label{eq:hllimit}
	\end{equation}
	where $N$ is the number of $^{136}$Xe atoms, $T$ is the live time, $\epsilon_{s}$ is the signal efficiency, and $S_{up}$ is the upper limit on the number of (NL)DBD-ES signal events at 90\% C.L.

The signal efficiency includes efficiencies from energy and topological cuts.
For experimental searches with a large number of backgrounds $B$, $S_{up} = 1.64\sqrt{B}$, where $\sqrt{B}$ denotes the Poissonian fluctuation of background.
Therefore, the half-life limit is proportional to $\epsilon_{s}/\sqrt{\epsilon_{b}}$, where $\epsilon_{b}$ is the efficiency that background events are kept by selection cuts.
Then, we use $\epsilon_{s}/\sqrt{\epsilon_{b}}$ as a figure of merit to optimize our topological cuts.
We used the Feldman--Cousins approach~\cite{FC} for calculating $S_{up}$ for searches with a small number of backgrounds (less than 25 counts), which takes care of the statistical fluctuation of small values.

\section{Double beta decay to $ 0_{1}^{+} $ excited state}
\label{sc:bb2n_Ex}

We define the analysis region of interest (ROI) for DBD to $0_{1}^{+}$ excited state based on main tracks (MT), which is the most energetic track in an event.
The sum of the energy of MT in the X--Z and Y--Z planes is called \emph{MTE}.
In the MTE spectra, ROIs of [725.7, 789.0] and [789.0, 854.6]~keV are defined around two de-excitation $ \gamma $ peaks at 760.5 and 818.5~keV.
The range of the ROIs is 2$ \sigma $ from the center, but the widths of the two connecting sides are shrunk to 1.6$ \sigma $ to avoid overlapping.
Eight additional parameters are defined based on the characteristics of reconstructed tracks, described as follows.

Three global parameters are the total deposited energy (denoted as $E_t$), the total number of tracks (\emph{$N_{t}$}), and the track dispersion ($ \emph{K}$), the last of which is defined as follows:
\begin{equation}
K^{XZ(YZ)}= \sum_{i=1}^{N} E_{i}  \quad \rm{for\, hits }\in S,
\end{equation}
where $E_{i}$ represents the energy of the $ i $th hits of an event within a circle $ S $, which is centered at the energy-weighted center of the event on the X--Z (Y--Z) plane.
The radius of $S$ is optimized to be 12.5~mm.

Three more parameters related to MT are the projected MT length at each plane (\emph{L}), a ratio of the total energy of all sub-dominant tracks to the total deposited energy of the event (\emph{R = 1 $-$ MTE/$E_t$}), and blob ratio of the MT ($ \emph{QR}_{1} $).
Furthermore, we introduce the blob ratio of the second most energetic track ($ \emph{QR}_{2} $).

The last parameter $\Delta$ aims to identify the de-excitation $\gamma$ rays, even if the energy deposition occurs in more than one track.
The definition of $\Delta$ follows EXO-200~\cite{EXO-200},
\begin{equation}
\Delta_{i} =\min_{j} (|E_{j}-\gamma_{i}|),
\end{equation}
where $i$ represents the two cases where $ \gamma_{1}\, (\gamma_{2})=760.5\,(818.5)$~keV. $ E_{j} $ iterates all possible combinations of an event’s track energy.

Figure~\ref{TMVA_DBD-ES} illustrates the distributions of parameters in MTE ROI [725.66, 788.96]~keV.
Length cuts $ L^{XZ}\neq 0 $ and $L^{YZ} \neq 0 $ are used to remove very short $ \alpha $ particle tracks originating from the Micromegas' surface for all plots in the figure.
For the projection plane specific parameters, only those for the X--Z plane are shown.
In this MTE ROI, the signal event may include additional electrons and/or $\gamma$ tracks in addition to the 760.5~keV $\gamma$.
Most of the background comprises single $\gamma$s.
The distinction is demonstrated in the $E_t$ and $N_t$ distributions.
Because of the dispersion of signals, the K value of the signals is close to zero.
The more centered distribution of signals' $\Delta$ value in the top right panel is attributed to its particular $\gamma$'s energy.
Furthermore, since the events are selected based on MTE, the distributions of the MT length are comparable for signal and background.
$R$ of the background has a prominent peak at zero because most of the background events have only one track at X--Z (Y--Z) plane, while the distribution of $R$ of signal is dominated by the energy of two electrons over $E_t$.
In the case of only one track in background events, QR$_{2} $ is not defined and assigned to be $-$1.
However, since the secondary tracks can be short and the blobs at the ends cover the entire track, a large fraction of the background events have QR$_{2} $ equal to 1.
As the secondary tracks are from two electrons of DBD, the QR$_{2} $ of the signal peak is close to one
\begin{figure} [H]
    	\centering
    	\includegraphics[width=0.485\textwidth]{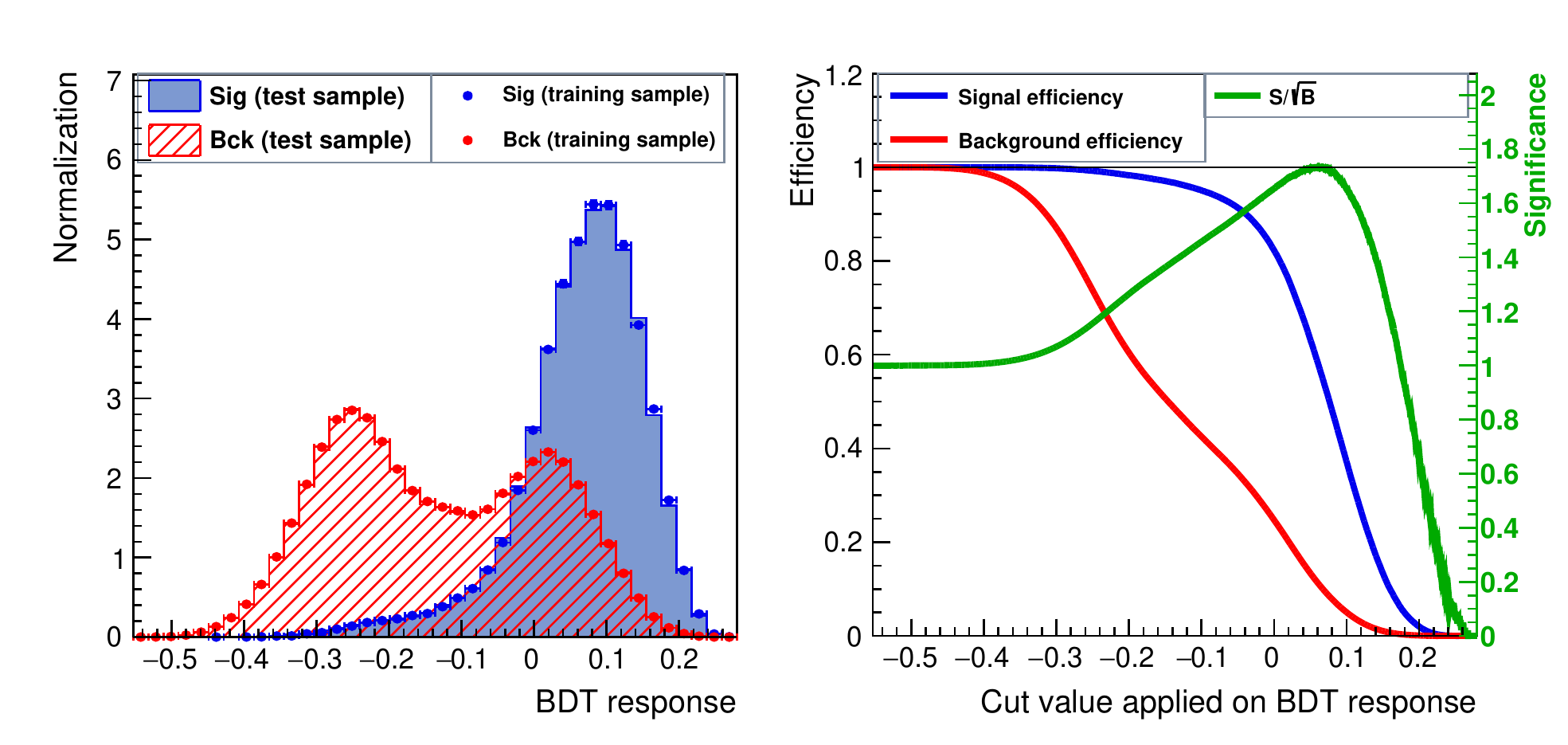}
    	\caption{Distribution of TMVA discriminator in [725.7, 789.0]~keV MTE ROI of DBD-ES (left) and signal efficiency, background efficiency, and discrimination significance versus the BDT response cut (right). }
    	\label{DBDCurve}
\end{figure}

As previously stated, TMVA tags each event with a discriminator value between $-$1 (background-like) and 1 (signal-like.
We determine the cut value by maximizing the signal-to-background significance $ \epsilon_{s}/\sqrt\epsilon_{b} $, where $ \epsilon_{s} $ and $ \epsilon_{b} $ represent the efficiency of signal and background selection after the cut, respectively.
We have investigated the discriminating power of different algorithms, including the boosted decision tree (BDT), boosting decision tree gradient (BDTG), support vector machines (SVM), and likelihood.
Among them, BDT provides the best discriminating power, although the difference is small.
In this study, the BDT approach is exclusively used.
Figure~\ref{DBDCurve} (left) illustrates the distribution of the TMVA discriminator for background and signal in $[725.7, 789.0]$~keV MTE ROI.
The background discriminator distribution has two peaks that are related to the number of tracks in an event.
In particular, background events with a single track are more likely to have smaller discriminator values, and background events with multiple tracks are categorized as more signal-like.
Figure~\ref{DBDCurve} (right) shows the efficiencies and significance as a function of the discriminator.
When the BDT cut is set at 0.061, the best discrimination significance of 1.7 is reached.
\begin{figure}[H]
	\includegraphics[width=0.485\textwidth]{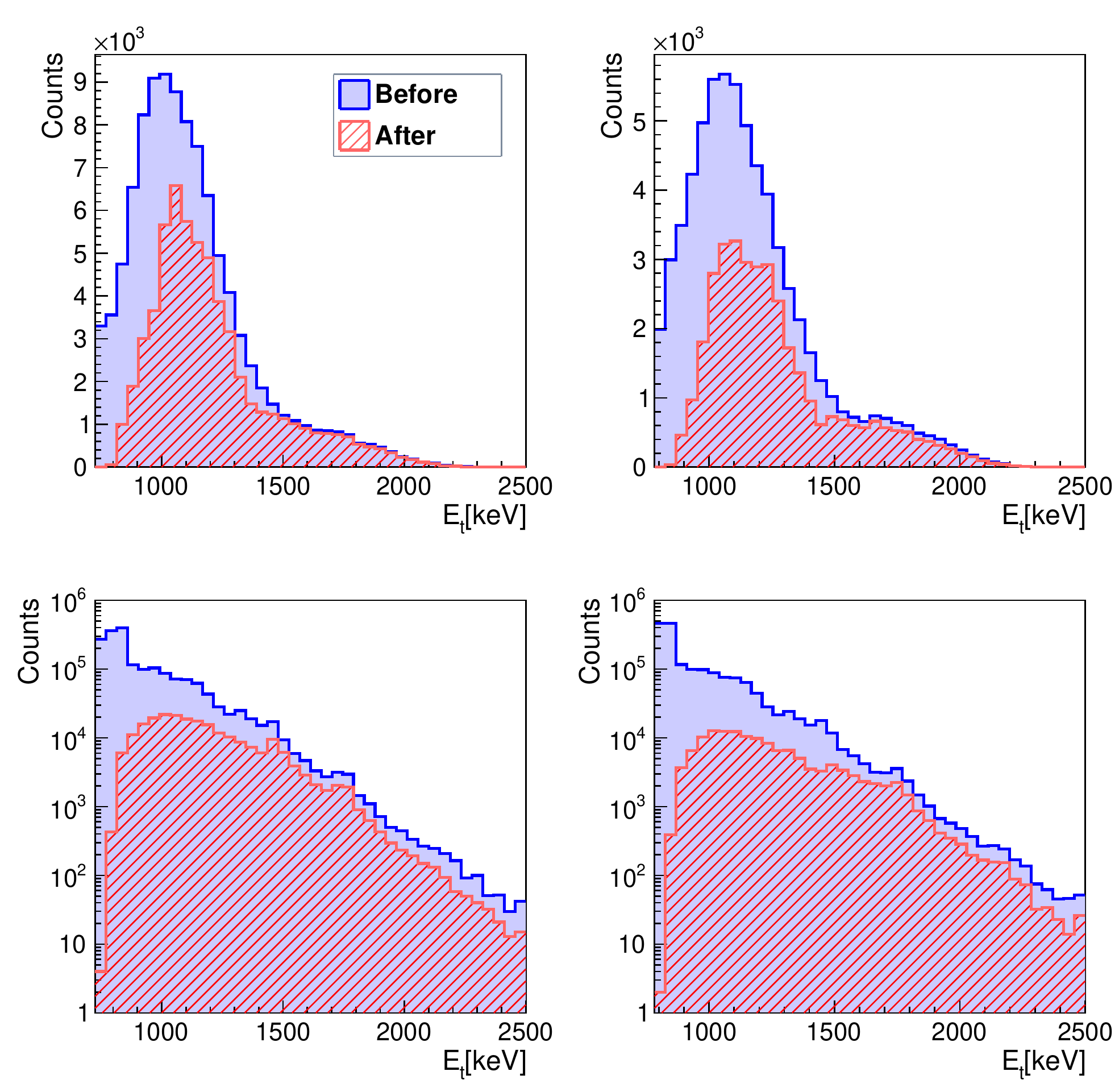}
	\caption{Effects of topological cuts for the DBD-ES signal (top two) and background (bottom two). The left column is for MTE ROI of 760.5~keV, and the right column is for 818.5~keV. The signals (backgrounds) before topological cuts are shown in the solid blue line, while those after topological cuts are shown in the solid red line. Note that signal spectra are in linear scale and background spectra are in logarithmic scale.
}
	\label{DBD-ESCut}
\end{figure}
Figure~\ref{DBD-ESCut} shows the energy spectra of signal and background events before and after topological cuts for the two MTE ROIs.
\begin{table}[H]
	\centering
	\renewcommand\arraystretch{1.5}
	\resizebox{\columnwidth}{!}
{
            \begin{tabular}{cccccc}
             \toprule[0.2mm]
			 \multirow{2}{*}{MTE ROI [keV]}	 &\multicolumn{3}{c}{Significance}	 & \multicolumn{2}{c}{Efficiency}  \\
                              &  MTE &	Topology & Total & Signal & Background \\
			\midrule[0.1mm]
			{[725.7, 789.0]} & 3.3 &	1.7 &	5.7	&	1.2\%	& 4.4$\times$10$^{-6}$
			\\
			{[789.0, 854.6]} &2.3 &	1.9	&	4.3	&	 0.7\%	& 2.7$\times$10$^{-6}$
			\\
            {[725.7, 854.6]} & 3.9 &	1.8	&	7.1	&1.9\% &	7.1$\times$10$^{-6}$
			\\
			\bottomrule[0.2mm]

		    \end{tabular}
	}
	\caption{Significances of MTE, topology, and total cuts in the two MTE ROIs as well as the combined ROI.
	The last two columns list the total cut efficiencies of signal and background.
}
\label{table:DBD-significance}
\end{table}
\begin{figure*}[t]
	\centering
	\includegraphics[width=\linewidth]{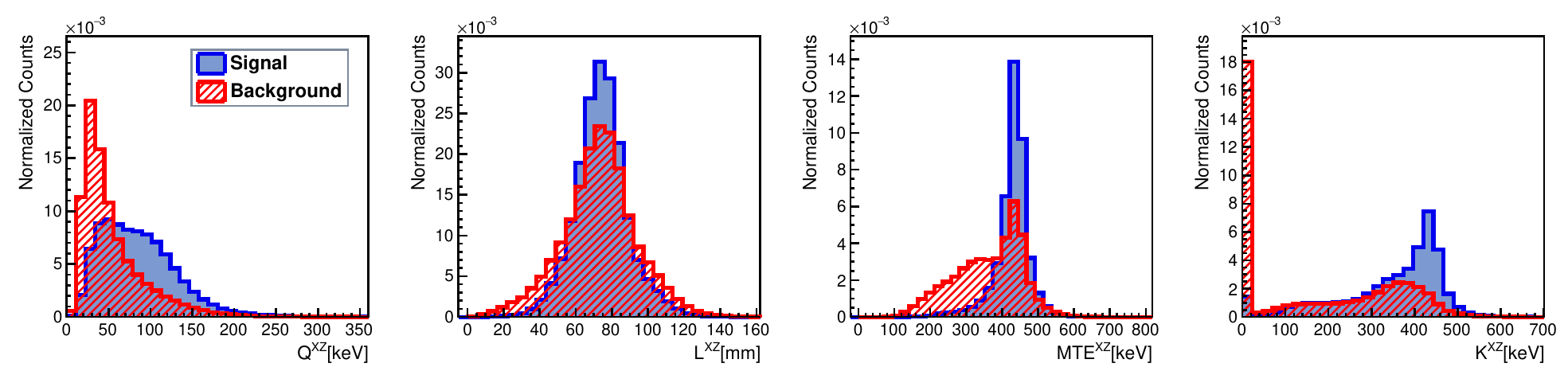}
	\caption{Distributions of the input topological parameters of TMVA in the 878.8$\pm$37.4~keV ROI. Signal (NLDBD-ES) is denoted in solid blue and background in hatched red.}
	\label{TMVA_NLDBD-ES}
\end{figure*}

\begin{figure*}[t]
	\centering
	\includegraphics[width=\linewidth]{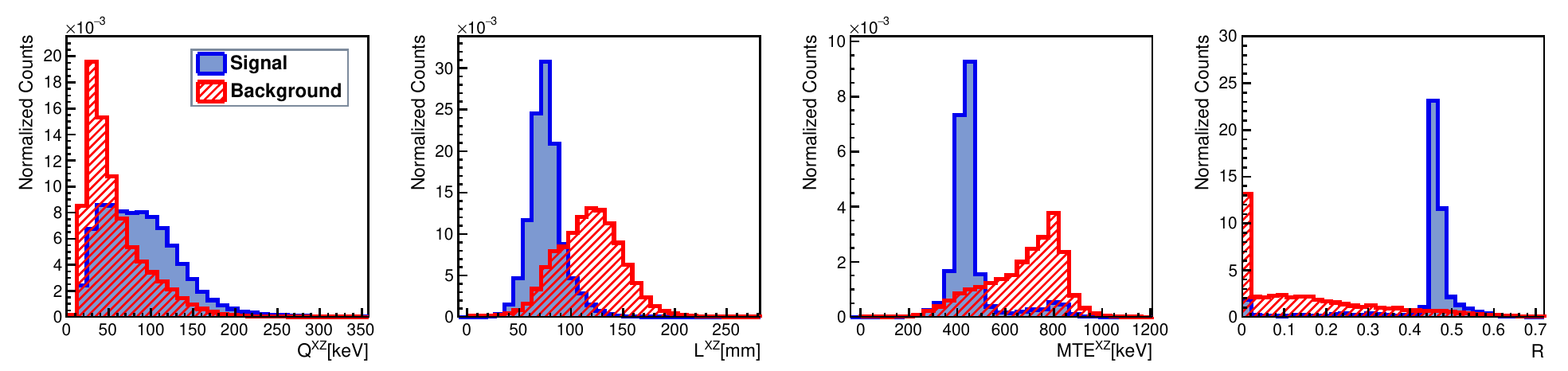}
	\caption{Distributions of the input topological parameters of TMVA in the 1639.3$\pm$51.1~keV ROI. Signal (NLDBD-ES) is denoted in solid blue and background in hatched red.}
	\label{TMVA_NLDBD-ES_1639}
\end{figure*}

Table~\ref{table:DBD-significance} shows the significance of the MTE ROI cut, topological cut, and combination of the two.
Note that topological significance is the product of the significances of length cuts and the TMVA cut.
We also list the total reduction efficiencies of signal and background.
The best significance is obtained when the two MTE ROIs are merged.
The total background can be reduced by more than five orders of magnitude while maintaining the DBD-ES signal efficiency at 1.9\%.
For topological cuts only, the signal (background) efficiency is 55.9\% (9.6\%).

After the cuts in TMVA are finalized, the improvement in DBD-ES half-life search sensitivity can be calculated.
Table~\ref{table:DBD-ES} shows the number of background counts from different sources in different MTE ROIs after three years of exposure with the PandaX-III detector.
With a total number of background events of 1779 in the merged ROI, the search sensitivity for DBD-ES is $4.1\times10^{23} $yr at the 90\% C.L., representing an improvement by a factor of 1.8 compared to that of no topological cuts.


\begin{table*}[t]
	\centering
	\renewcommand\arraystretch{1.5}
	\resizebox{2\columnwidth}{!}
	{
            \begin{tabular}{ccccccccccccc}
             \toprule[0.2mm]
			 \multirow{2}{*}{MTE ROI [keV]}	 &\multicolumn{4}{c}{ Liner Background}	 &\multicolumn{2}{c}{ MM Background} &  \multicolumn{2}{c}{ SS Background} & \multicolumn{2}{c}{ Acrylic Background} &  \multirow{2}{*}{\shortstack{Total \\{Background}}}& \multirow{2}{*}{\shortstack{Sensitivity \\{[yr]}}}
             \\   \cmidrule(r){2-5} \cmidrule(r){6-7} \cmidrule(r){8-9} \cmidrule(r){10-11}
                             &	$^{238}$U	& $^{232}$Th	&	$^{60}$Co & $^{40}$K & $^{238}$U & $^{232}$Th & $^{238}$U & $^{232}$Th & $^{238}$U & $^{232}$Th &    &
             \\
            \midrule[0.1mm]
            {[725.7, 789.0]} &	36	& 8	&	260 & 152 &115 & 30 & 20 & 17 &   378 & 94 & 110 & {3.3$\times$10$^{23}$}
            \\
           {[789.0, 854.6]} &	22	& 4	&	166 & 90 &75 & 15 & 12 & 12 &   228 & 45 & 669 & {2.5$\times$10$^{23}$}
           \\
           {[725.7, 854.6]} &	58	& 12	&	426 & 242 &190 & 45 & 32 & 29 &   606 & 139 & 1779 & {4.1$\times$10$^{23}$}
           \\
			\bottomrule[0.2mm]
		\end{tabular}
	}
		\caption{Counts in different MTE ROIs from each background source, assuming three years of exposure in PandaX-III.
		The last row lists the projected sensitivities at 90\% C.L.
}
	\label{table:DBD-ES}
\end{table*}

\section{Neutrinoless double beta decay to $ 0_{1}^{+} $ excited state}
\label{sc:bb0n_Ex}

Four possible ROIs were identified to search for NLDBD-ES depending on whether de-excitation $\gamma$ rays deposit their full energy in the active volume of TPC.
In all cases, the energy of two electrons is fully contained and recorded.
The four ROIs are $878.8\pm37.4$, $1639.3\pm51.1$, $1697.3\pm52.0$, and $2458.8\pm62.6$~keV, corresponding to events with no $\gamma$, one 760.5~keV $\gamma$, one 818.5~keV $\gamma$, and both $\gamma$ rays captured.
ROI ranges are defined as $\pm2\sigma$ around the center value, where $\sigma$ is the energy resolution, scaled from an expected energy resolution of $\sigma=1.3\%$ at $Q_{NLDBD}=2457.8$~keV for PandaX-III.
We will use the center value to denote each ROI later in the text.

We define five parameters specific to the topological signatures of NLDBD-ES, including the track dispersion (K) parameter, which is calculated using all possible tracks of an event and four parameters specific to the main track.
The MT-related parameters are MTE, L, R, and Q, which is the smaller of the two end blob energies along MT.
In the 878.8~keV ROI, we use all parameters except R.
For the other three ROIs, we only use MT-related parameters as input variables for TMVA.
The MT in signal events is the track of two electrons in all four ROIs since the energy of two electrons is larger than that of the other two $\gamma$s in NLDBD-ES.
Figure~\ref{TMVA_NLDBD-ES} and Figure~\ref{TMVA_NLDBD-ES_1639} show the distributions of each input topological parameters used in the 878.8 and 1639.3~keV ROIs, respectively.
The distributions of values in the X--Z plane are used for all parameters except R.
The distributions of Q for signals in both figures are similar since they are for the two-electron tracks.
Following that, the distributions of the other two MT parameters, L and MTE of signal, are almost the same in both ROIs.
The small peak near 800~keV in the MTE distributions in Figure~\ref{TMVA_NLDBD-ES_1639}
\begin{figure} [H]
    	\centering
    	\includegraphics[width=0.485\textwidth]{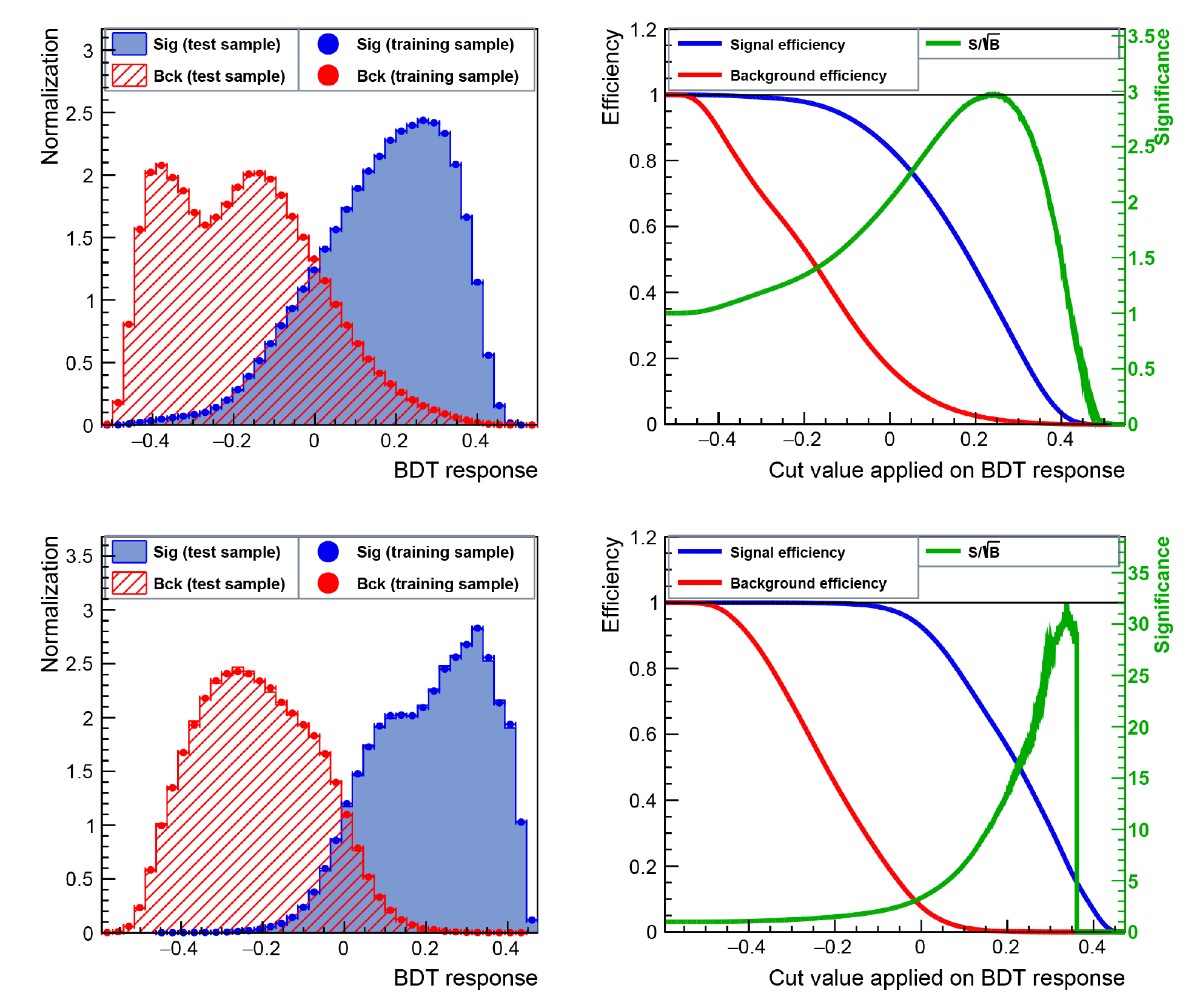}
    	\caption{The left two panels show the distributions of the TMVA discriminator in the 878.8~keV (top) and 1639.3~keV (bottom) ROIs of NLDBD-ES. The right panels show the corresponding signal efficiency, background efficiency, and significance for different BDT response cuts. The optimal significance is 3.0 (32.1) when cut at 0.234 (0.338) for the 878.8~keV (1639.3~keV) ROI.}
    	\label{NLDBD_Curve}
    \end{figure}

 \noindent is due to wrongly connected electron and $\gamma$ ray tracks.

Figure~\ref{NLDBD_Curve} shows the distribution of the TMVA discriminator and related efficiencies as well as significance in the 878.8 and 1639.3~keV ROI.
In the 878.8~keV ROI, two peaks of the background discriminator distribution are related to the number of tracks, similar to the DBD case (Section~\ref{sc:bb2n_Ex}).
Due to background fluctuations, the 1639.3~keV significance curve is cut off at the right end, where the number of background events is less than ten, and the calculated significance is no longer reliable.

In the 878.8~keV ROI, compared to the background events, NLDBD-ES events have larger Q, more concentrated MTE, and nonzero K.
All the characteristics are because signals are two-electron tracks.
In the 1639.3~keV ROI, most of the background events have a main track with high energy, thus affording long L, large MTE, and small or even zero R.
The distribution of R of signals in the last panel is centered at approximately 0.46, which is expected from the definition.

\begin{figure*}[t]
	\centering
	\includegraphics[width=\linewidth]{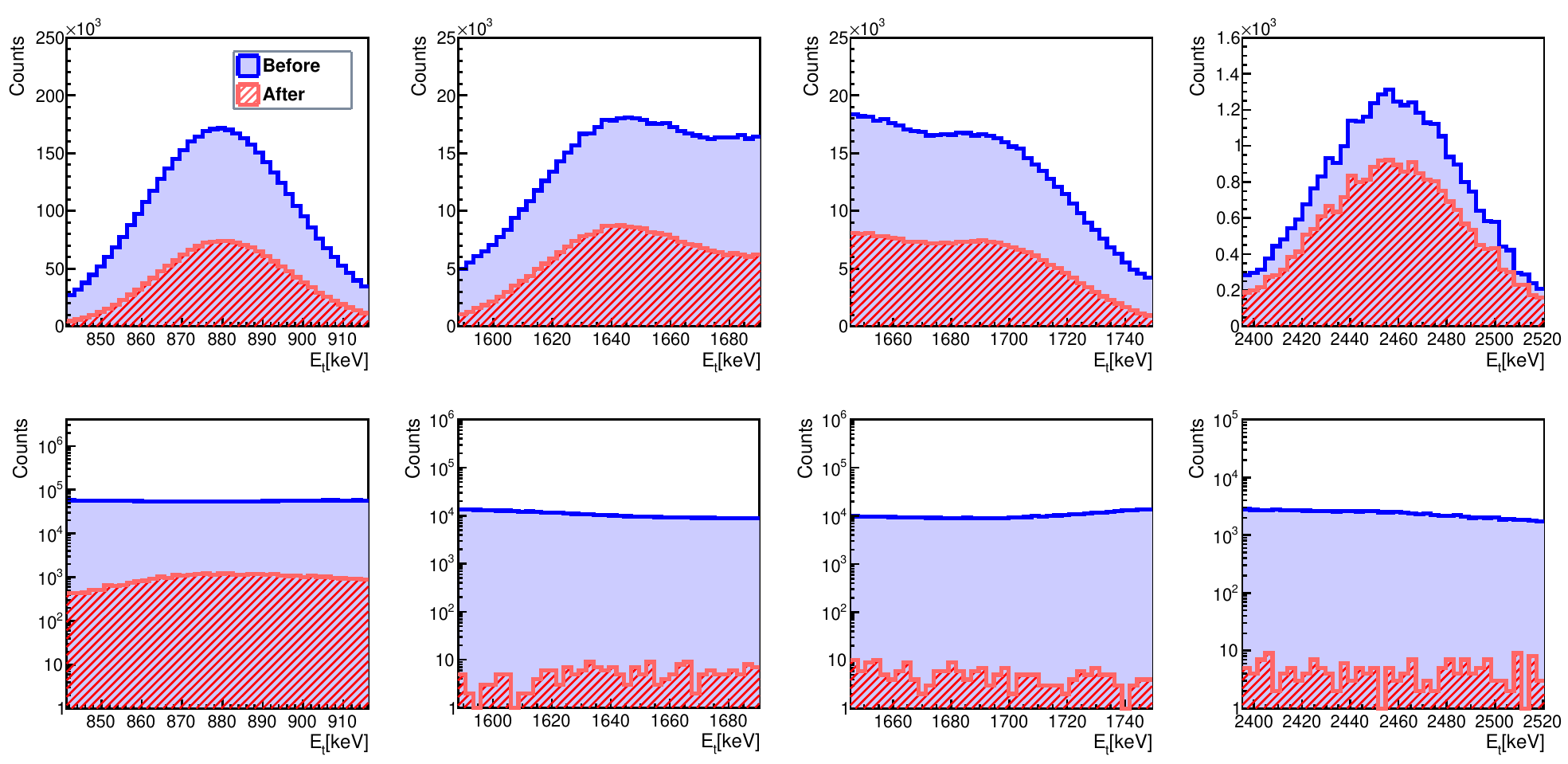}
	\caption{Effects of topological cuts for the NLDBD-ES signal (top) and background (bottom).
	The signal (background) before topological cuts is shown in blue, while that after topological cuts is shown in red.
	From left to right, each column represents the energy ROI centered at 878.8, 1639.3, 1697.3, and 2457.8~keV, respectively.
	Note that signal spectra are shown in linear scale and the background are shown in logarithmic scale.}
	\label{NLDBD-ESCut}
\end{figure*}

\begin{table}[H]\tiny
	\centering
	\renewcommand\arraystretch{1.5}
	\resizebox{\columnwidth}{!}
	{

             \begin{tabular}{cccc}
             \toprule[0.2mm]
			 \multirow{2}{*}{ ROI [keV]}	 & Significance	 & \multicolumn{2}{c}{Efficiency}  \\
                                                                             &	Topology & Signal & Background \\
			\midrule[0.1mm]
			{878.8} & 3.0	&	39.0\%	& 1.7$\times$10$^{-2}$
			\\
			{1639.3} &	32.6	&	 21.3\%	& 4.3$\times$10$^{-5}$
			\\
            {1697.3} &	28.7	&21.1\% &	5.4$\times$10$^{-5}$
			\\
            {2457.8} &	46.5	&47.7\% &	1.1$\times$10$^{-4}$
            \\
			\bottomrule[0.2mm]

		\end{tabular}
	}
	\caption{Topological significances and related reduction efficiencies (last two columns) of signal and background in each ROI.
}
	\label{table:NLDBD-significance}
\end{table}

\begin{table*}[t]
	\centering
	\renewcommand\arraystretch{1.5}
	\resizebox{2\columnwidth}{!}
	{

        \begin{tabular}{cccccccccccccc}
             \toprule[0.2mm]
			 \multirow{2}{*}{ROI [keV]}	 &\multicolumn{4}{c}{ Liner Background}	 &\multicolumn{2}{c}{ MM Background} &  \multicolumn{2}{c}{ SS Background} & \multicolumn{2}{c}{ Acrylic Background} & \multirow{2}{*}{\shortstack{Total \\{Background}}}& \multirow{2}{*}{\shortstack{Signal \\{Efficiency}}}& \multirow{2}{*}{\shortstack{Sensitivity \\{[yr]}}}
             \\ \cmidrule(r){2-5} \cmidrule(r){6-7} \cmidrule(r){8-9} \cmidrule(r){10-11}
                             &	$^{238}$U	& $^{232}$Th	&	$^{60}$Co & $^{40}$K & $^{238}$U & $^{232}$Th & $^{238}$U & $^{232}$Th &  $^{238}$U & $^{232}$Th &  &   &
             \\
            \midrule[0.1mm]
            {878.8} &	6.0	& 2.1	& 41.9 & 26.8 &29.0 & 6.5 & 3.4 & 3.3 &   47.7 & 20.0 & 186.6 & 16.3\% & {1.1$\times$10$^{25}$}
            \\
           {1639.3} &	0.0	& 0.0 & 0.2 & 0.0 & 0.2 & 0.0 & 0.0 & 0.0 &   0.4 & 0.1 & 1.0 & 2.4\% & {1.7$\times$10$^{25}$}
           \\
           {1697.3} &	0.0	& 0.0	&	0.2 & 0.0 &0.3 & 0.0 & 0.0 & 0.0 &   0.3 & 0.1 & 1.0& 2.3\% & {1.7$\times$10$^{25}$}
           \\
           {2457.8} &	0.0	& 0.0	&	0.3 & 0.0 &0.2 & 0.1 & 0.0 & 0.0 &   0.1 & 0.2 & 0.9 & 0.2\% & {1.5$\times$10$^{24}$}
           \\
			\bottomrule[0.2mm]
         \end{tabular}

	}
	\caption{Counts in different energy ROIs from each background source assuming three years of exposure in PandaX-III.
		The last two columns list the signal efficiencies and half-life sensitivities at 90\% C.L..}
	\label{table:NLDBD-ES}
\end{table*}

Table~\ref{table:NLDBD-significance} lists the efficiencies and significances of topological cuts in four different ROIs, and Figure~\ref{NLDBD-ESCut} displays the NLDBD-ES and background spectra before and after topological cuts. The background was reduced with topological cuts by 2--5 orders while maintaining the NLDBD-ES efficiency between $21\%$ and $48\%$ in different ROIs.
With additional $\gamma$ rays, the latter three ROIs' significances are better than the first one as well as those in the DBD-ES cases.

The background counts after energy and topological cuts of three years of exposure in PandaX-III are given in Table~\ref{table:NLDBD-ES}.
The cuts are so effective that we reject almost all the background events in the latter three ROIs.
Compared with Table~\ref{table:NLDBD-significance}, we relax the TMVA cut thresholds to retain more signals and thus obtain better sensitivity.
The final cut criteria are optimized to afford the best half-life search sensitivity with the Feldman--Cousins approach and Equation~\ref{eq:hllimit}. The corresponding signal efficiencies and search sensitivities are shown in the last two rows.
The best search sensitivity, 1.7$\times$10$^{25}$~yr (90\% C.L.), is given in the 1639.3~keV ROI, indicating improvement by a factor of 4.8 in comparison to that of no topological cuts. Here we do not merge ROIs as we do in the DBD-ES case, because there is an overlap between the two middle ROIs.

The first three ROIs afford a low limit of about 10$^{25}$~yr, while the magnitude of the last ROI is 10$^{24}$~yr.
For the 878.8~keV ROI, 41.9\% of the signals fall in this ROI, but the number of backgrounds is also the largest.
For the 2457.8~keV ROI, the topological cut is significant, but it suffers an extremely small signal efficiency due to the small probability of 0.3\% for both de-excitation $\gamma$s fully contained in the gas medium.

\section{Conclusion and discussion}
\label{sc:conclusion}

We have presented an improvement in search sensitivity for (NL)DBD-ES with the unique ability of gaseous detectors to record topological information of event trajectories.
Our study is based on a detailed simulation of the PandaX-III detector, with its expected detector performance and background budget.
We retain 1.9\% of the total number of DBD-ES signals while reducing the background by about five orders of magnitudes using MTE energy cut and topological characteristics.
With three years of PandaX-III data, the estimated sensitivity for DBD-ES reaches $ 4.1\times10^{23} $~yr (90$\%$ C.L.), representing a factor of 1.8 improvements compared to that without topological analysis.
With three years of exposure, the sensitivity of the $^{136}$Xe NLDBD-ES is $ 1.7 \times10^{25} $~yr (90$\%$ C.L.), corresponding to a signal efficiency of 21.3\% and background efficiency of $4.3\times 10^{-5}$ in the 1639.3~keV ROI.
Topological analysis improves the search sensitivity of NLDBD-ES by 4.8 times.

Further improvement can be expected in several aspects.
The search sensitivity of PandaX-III is limited by background for DBD-ES.
The current theoretical estimations of the half-life of DBD-ES range from $ 10^{23} $ to $ 10^{25} $ yr~\cite{Theory, EXO-200Theory}.
Further refinement of the TMVA algorithm parameters and new deep learning techniques improve the background suppression efficiency and thus the search sensitivity to cover more theoretical range.
For NLDBD-ES, the search sensitivity is limited by the number of signals.
The current best limit on the $^{136}$Xe NLDBD to $ 0_{1}^{+} $ $^{136}$Ba half-life is $2.4\times10^{25}$ yr (90\% CL), given by KamLAND-Zen~\cite{KamLAND-Zen:2nbb-ES}.
A larger gaseous detector would be necessary to further improve the search sensitivity beyond the current limit.

\section{Acknowledgements}
\label{sc:acknowledgements}
We thank Dr.~Damien Neyret and Dr.~Yann Bedfer for the fruitful discussion. This work is supported by the grant from the Ministry of Science and Technology of China (No.~2016YFA0400302) and the grants from National Natural Sciences Foundation of China (No.~11775142 and No.~11905127). This work is supported in part by the Chinese Academy of Sciences Center for Excellence in Particle Physics (CCEPP).

\end{multicols}

\end{document}